\documentclass[12p]{article}
\usepackage{latexsym}
\usepackage{color}
\usepackage{amsmath}
\usepackage{epsfig}
\usepackage{hyperref}
 \usepackage{dcolumn}
\newcommand{\ortala}[1]{\begin{center}#1\end{center}}

\newcommand{\sandd}[1]{\left\langle #1\right\rangle}

\newcommand{\summ}[3]{{{\underset{#1 }{\overset{#2}{\displaystyle\sum}}}#3}}

\newcommand{\re}[1]{(\ref{#1})}

\newcommand{\eq}[2]{\begin{equation}\label{#1}  #2\end{equation}}

\newcommand{\paran}[1]{\left(#1\right)}

\newcommand{\sch}[1]{Schrodinger}

\newcommand{\komb}[2]{\paran{\begin{array}{c} #1 \\ #2 \end{array}}}

\linethickness{0.6pt}

\begin{document}


\ortala{\textbf{Magnetic properties of the multi-walled nanotubes constituted by localized magnetic moments: spin-1/2 case}}

\ortala{\textbf{\"Umit Ak\i nc\i \footnote{umit.akinci@deu.edu.tr}}}

\ortala{\textit{Department of Physics, Dokuz Eyl\"ul University,
TR-35160 Izmir, Turkey}}

\section{Abstract}

Magnetic properties of the multi-walled nanotubes have been investigated. Heisenberg model, which is a suitable model
for the system consist of atoms with localized wave functions has been used.  Effective field theory in two spin cluster for spin-1/2
has been numerically solved and critical properties such as phase diagrams as well as the behavior 
of the order parameter and hysteresis loops obtained.  

Keywords: \textbf{ Multi-walled nanotube; Anisotropic Heisenberg model;
hysteresis loops }

\section{Introduction}\label{introduction}

Nanotubes are the promising materials and paid attention to these systems especially after the discovery of carbon nanotubes
\cite{ref1}, in both theoretically and experimentally. Since that day many 
types 
of nanotubes fabricated as well as ferromagnetic
nanotubes \cite{ref2,ref3}. These materials are promising materials for various 
technological applications \cite{ref4,ref5}.
Experimental works followed by determination of the magnetic properties with some theoretical methods, such as
mean field approximation (MFA),  effective field theory (EFT) and Monte Carlo simulation techniques (MC). 
Various nanotubes, such as $FePt$ and $Fe_3O_4$ nanotubes    \cite{ref6},
can be modeled by core-shell models. Handling theoretical solutions within the core-shell models widely adopted in the literature,
starting from the most basic model of the statistical physics namely Ising model 
\cite{ref7}. 

It is a well-known fact that magnetic order is originated from the exchange interaction between atoms that have magnetic
dipole moments, come from the spin magnetic moment. Models can be grouped into two classes namely, related to the magnetism comes from the localized moments and related to the itinerant magnetism. Itinerant magnetism occurs mostly in transition metal alloys and comes from the filled delocalized states by electrons. Hubbard model
\cite{Hubbard} is the well-suited model in this case. On the other hand if the wave 
functions 
are don't overlap significantly each other, magnetism comes from the localized moments. In other words, if atoms of the system are
well separated, and wave functions are localized, then we can say that we are far from itinerant magnetism and using 
Heisenberg model is appropriate. Although Heisenberg model is the general model, its anisotropic limit Ising model on the core-shell structured nanotube geometry is the most widely studied in the name of the magnetic properties
of the nanotubes.  There has been less attention paid on the Heisenberg model 
on 
the nanotube geometry within the core-shell models.
Classical Heisenberg model on a
single wall ferromagnetic nanotubes has been solved with MC \cite{ref8,ref9} 
and 
many-body Green's function method \cite{ref10}.
Also slightly different geometry, namely three-leg quantum spin
tube \cite{ref11} has been solved with numerical exact diagonalization within 
the finite-cluster \cite{ref12}.

Core-shell treatments have tubular geometry but contain only two interacted cylindrical structures. On the other hand, this class of treatment covers a tiny part of the nanotubes.
Thus treatments with more 
general geometries necessary such as multi-walled 
nanotubes (MWNT). What is the effect of the number of walls of the nanotube on the magnetic properties of it? This is the central
question of this work. Phase diagrams, thermodynamical quantities 
(magnetization, magnetic susceptibility) and hysteresis
characteristics will be obtained for MWNT. As a formulation, we use the EFT 
formulation in a 2-spin cluster (EFT-2). 
EFT  can provide results that are superior to those obtained within
the MFA, due to the consideration of self-spin correlations, which are omitted 
in the MFA.
EFT-2 formulation \cite{ref13} mostly used for the Heisenberg model. 

The paper is organized as follows: In Sec. \ref{formulation} we
briefly present the model and formulation. The results and
discussions are presented in Sec. \ref{results}, and finally Sec.
\ref{conclusion} contains our conclusions.

\section{Model and Formulation}\label{formulation}

\begin{figure}[h]\begin{center}
\epsfig{file=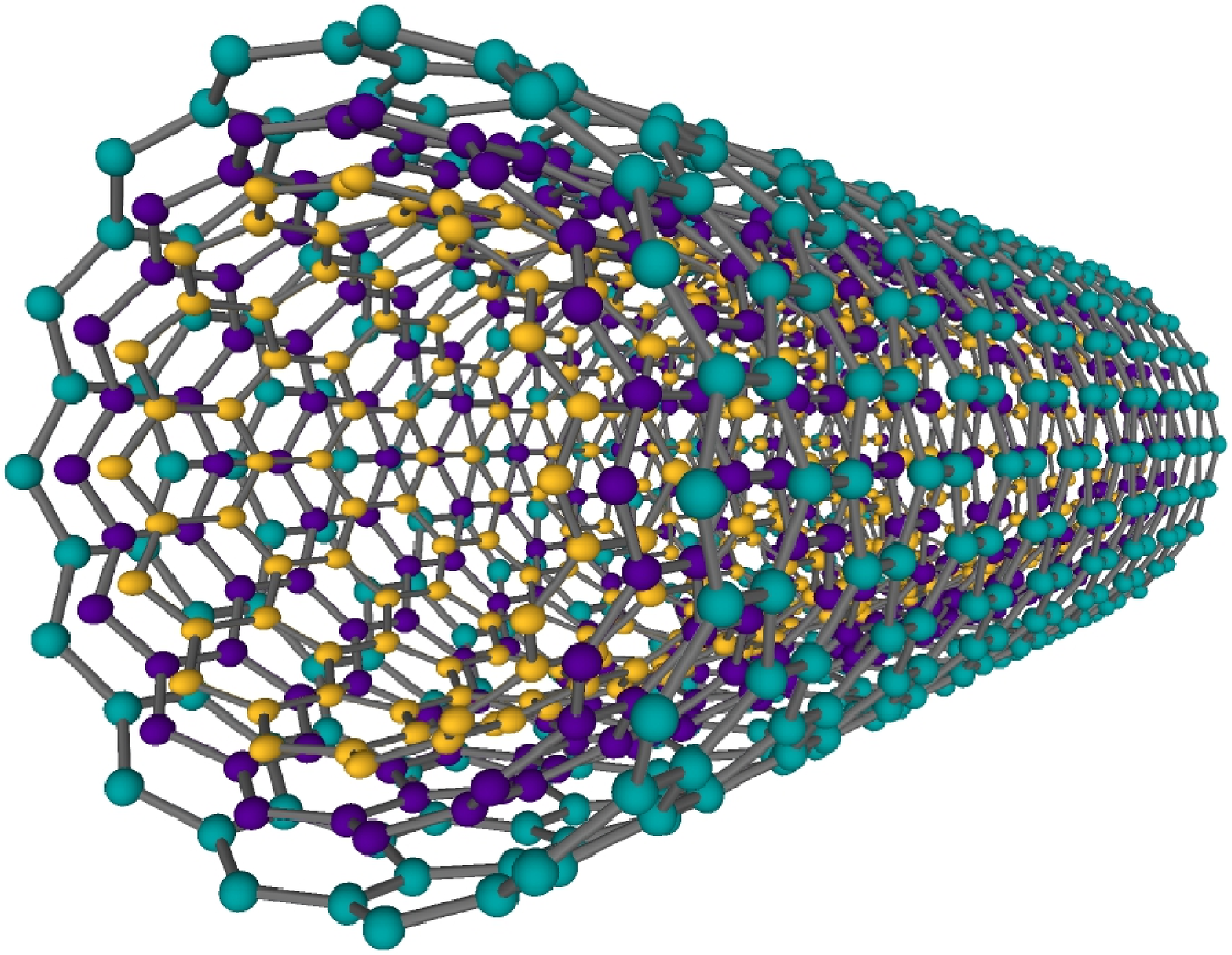, width=7.2cm}
\epsfig{file=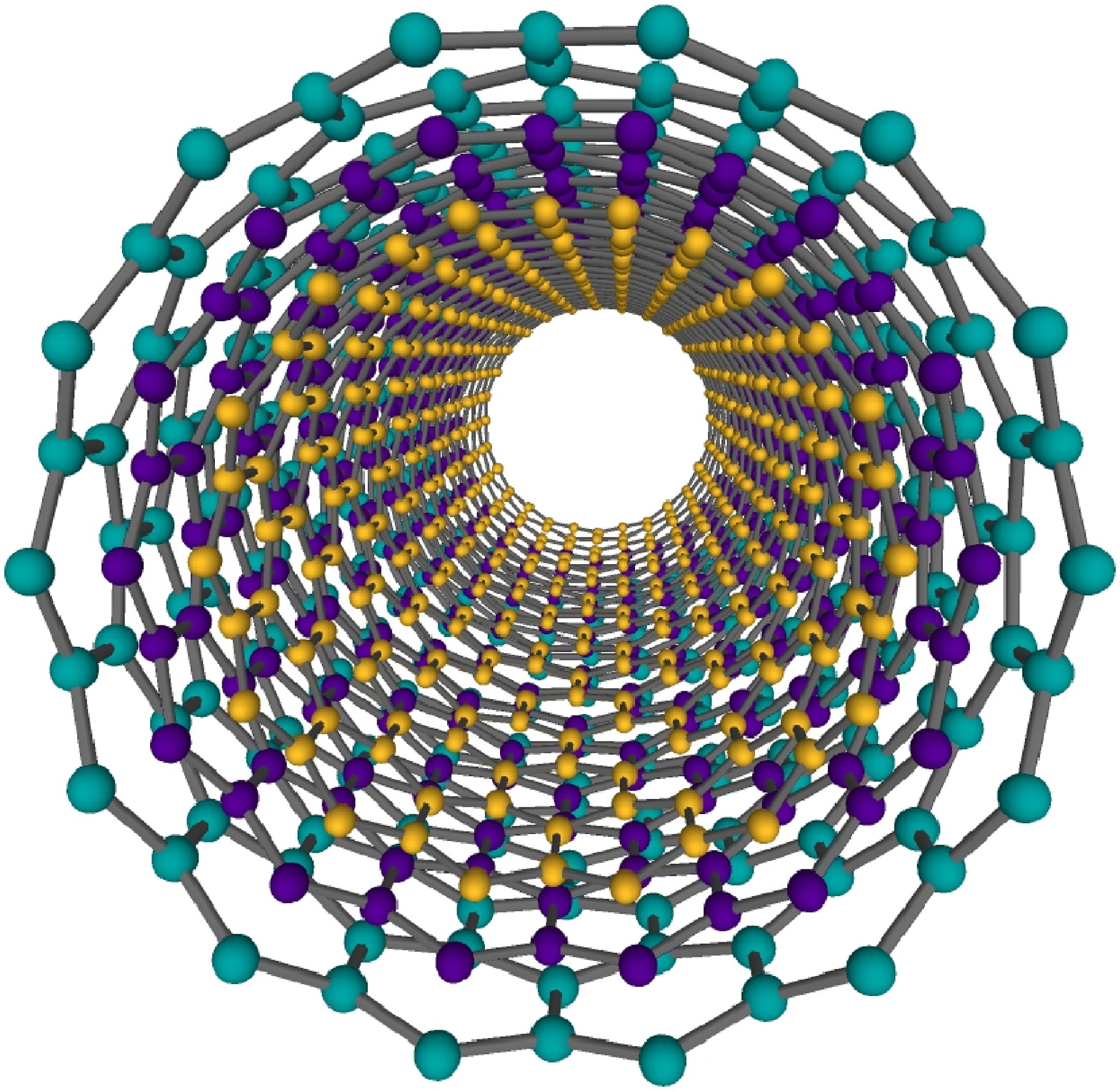, width=7.2cm}
\end{center}
\caption{Schematic representation of the system} \label{sek1}\end{figure}

The schematic representation of the system can be seen in Fig. \ref{sek1} in 
two 
different perspective. The system consist
of a nested number of $L$ interacted tubular system which all of them contain hexagonal lattice in it.

The Hamiltonian of the spin-1/2 anisotropic Heisenberg model on MWNT geometry 
is 
given by

\eq{denk1}{\mathcal{H}=-\summ{<i,j>}{}{J_{ij}\left[\Delta_{ij}\paran{ 
S_i^xS_j^x+S_i^yS_j^y}+ S_i^zS_j^z\right]}-
H\summ{i}{}{S_i^z}}
where $S_i^x,S_i^y$ and  $S_i^z$ denote the Pauli spin-1/2 operators at a site 
$i$. $J_{ij}$ and $\Delta_{ij}$  are the
exchange interaction and anisotropy in the exchange interactions between the 
nearest
neighbor spins located at sites $i$ and $j$, respectively.
In our system, all spins are under the influence of the longitudinal magnetic field $H$. In Eq. \re{denk1}
the first sum is carried over the nearest neighbors of the lattice, while the second one is overall the lattice sites.

EFT with finite cluster has a simple strategy: constructing finite cluster which spin-spin interactions in the cluster
treated exactly and making an approximation with using local fields to
treat interaction between the cluster and outside of it. 
With using exact spin identities such as Callen-Suzuki identities 
\cite{ref14,ref15} or generalization of them \cite{ref16}
we can calculate any observable of the system. Due to the avoid mathematical 
difficulties, 
the interaction between the cluster and outside of it  treated within the Ising 
type 
interaction namely axial approximation \cite{ref17}.
To include Heisenberg character of the interaction to the formulation, minimum cluster size choice should be two. Then we use
EFT-2 here \cite{ref18}. In this approximation we assume also the translational 
invariance of the $i^{th}$ cylinder, i.e. in $i^{th}$
cylinder all sites are equivalent. Under these assumptions magnetization of the 
$i^{th}$ cylinder can be calculated via

\eq{denk2}{
m_i=\sandd{S_i^z}=\sandd{\frac{Tr_{2}S_i^z\exp{\paran{-\beta 
\mathcal{H}_i^{(2)}}}}{Tr_{2}
\exp{\paran{-\beta \mathcal{H}_i^{(2)}}}}}, \quad i=1,2,\ldots, L
}
Here, $Tr_2$ stands for the trace operation over the spins which are
belong to the chosen cluster and $\mathcal{H}_i^{(2)}$ is the finite cluster 
Hamiltonian constructed in $i^{th}$ cylinder:
\eq{denk3}{
-\beta \mathcal{H}_i^{(2)}=\beta J_i\left[\Delta_i\paran{ 
S_{i,1}^xS_{i,2}^x+S_{i,1}^yS_{i,2}^y}+S_{i,1}^zS_{i,2}^z\right]+\beta
\paran{H+h_1^{(i)}}S_{i,1}^z+\beta\paran{H+h_2^{(i)}}S_{i,2}^z
} where $i=1,2,\ldots,L$ and $J_i,\Delta_i$ are the intralayer exchange 
interaction and anisotropy in the exchange interaction about
the $i^{th}$ cylinder. $\beta=1/(k_BT)$, where $k_B$ is Boltzmann constant and 
$T$ is the absolute temperature. In finite cluster Hamiltonians $S_{i,1}, 
S_{i,2}$ are the spins belong to the chosen cluster 
in $i^{th}$ cylinder, and $h_1^{(i)},h_2^{(i)}$ are stand for the interaction
of these spins with other spins which are outside of the cluster. Let 
$S_{i,3},S_{i,4}$ are the nearest neighbor spins
(which are outside of the cluster) of the $S_{i,1}$ and  $S_{i,5},S_{i,6}$ are 
the nearest neighbor spins
of the $S_{i,2}$. 
Let the pair $(J_{1},\Delta_{1})$  be the exchange interaction and the 
anisotropy in the exchange interaction related to
the outer and inner cylinder, i.e. the nearest neighbor spins which are in the 
interacted via these couple. Let all  the 
other interactions given by $(J_{2},\Delta_{2})$.  Then interactions with the 
outside of the cluster are given as 
\eq{denk4}{
\begin{array}{lcl}
h_1^{(1)}&=&J_1\paran{S_{1,3}^z+S_{1,4}^z}+J_2S_{2,1}^z\\
h_2^{(1)}&=&J_1\paran{S_{1,5}^z+S_{1,6}^z}+J_2S_{2,2}^z\\
h_1^{(k)}&=&J_2\paran{S_{k,3}^z+S_{k,4}^z+S_{k-1,1}^z+S_{k+1,1}^z}\\
h_2^{(k)}&=&J_2\paran{S_{k,5}^z+S_{k,6}^z+S_{k-1,2}^z+S_{k+1,2}^z}\\
h_1^{(L)}&=&J_1\paran{S_{L,3}^z+S_{L,4}^z}+J_2S_{L-1,1}^z\\
h_2^{(L)}&=&J_1\paran{S_{L,5}^z+S_{L,6}^z}+J_2S_{L-1,2}^z\\
\end{array}
}

In order to calculate Eq. \re{denk2} we need to calculate the matrix 
representation of finite cluster Hamiltonian
$\mathcal{H}_k^{(2)}$, then exponentiate it. After some algebra, expression for 
the order parameter of the system which is defined by
\eq{denk5}{
m_i=\frac{1}{2}\paran{\sandd{S_{i,1}^z}+\sandd{S_{i,2}^z}},}
can be written as
\eq{denk6}{
\begin{array}{lcl}
m_k&=&\sandd{\frac{\sinh{\paran{\beta X_0^{(k)}}}}{\cosh{\paran{\beta 
X_0^{(k)}}}+
\exp{\paran{-2\beta J_1}}\cosh{\paran{\beta Y_0^{(k)}}}}}\\
m_l&=&\sandd{\frac{\sinh{\paran{\beta X_0^{(l)}}}}{\cosh{\paran{\beta 
X_0^{(l)}}}+
\exp{\paran{-2\beta J_2}}\cosh{\paran{\beta Z_0^{(l)}}}}}\\
\end{array},
}
where 

\eq{denk7}{\begin{array}{lcl}
X_0^{(n)}&=&(h_1^{(n)}+h_2^{(n)}+2H),\quad n=1,2,\ldots,L\\
Y_0^{(k)}&=&\left[4J_1^2\Delta_1^2+(h_1^{(n)}-h_2^{(n)})^2\right]^{1/2}, \quad 
k=1,L\\
Z_0^{(l)}&=&\left[4J_2^2\Delta_2^2+(h_1^{(n)}-h_2^{(n)})^2\right]^{1/2}, \quad 
l=1,2,\ldots, L \\
\end{array}}where $\beta=1/(k_B T)$ where $k_B$ is Boltzmann
constant and $T$ is the temperature.

Magnetization expressions given in closed form in Eq. \re{denk6} can be 
calculated via differential operator technique with
decoupling approximation \cite{ref19}. Similar procedure can be found in  Ref. 
\cite{ref20}. With the help of the Binomial expansion, Eq. \re{denk6} can be 
written in the form
\eq{denk8}{\begin{array}{lcl}
m_1&=&\summ{p=0}{2}{}\summ{q=0}{2}{}\summ{r=0}{1}{}\summ{s=0}{1}{}K_1\paran{p,q,
r,s}m_1^{p+q} m_2^{r+s}\\
m_2&=&\summ{p=0}{1}{}\summ{q=0}{1}{}\summ{r=0}{3}{}\summ{s=0}{3}{}K_2\paran{p,q
,r,s}m_1^{p+q} m_2^{r+s}\\
m_k&=&\summ{p=0}{1}{}\summ{q=0}{1}{}\summ{r=0}{2}{}\summ{s=0}{2}{}\summ{t=0}{1}{
}\summ{v=0}{1}{}K_2\paran{p,q,r,s,t,v}m_{k-1}^{p+q} m_k^{r+s}m_{k+1}^{t+v}\\
m_{L-1}&=&\summ{p=0}{1}{}\summ{q=0}{1}{}\summ{r=0}{3}{}\summ{s=0}{3}{}K_2\paran
{p,q,r,s}m_L^{p+q} m_{L-1}^{r+s}\\
m_L&=&\summ{p=0}{2}{}\summ{q=0}{2}{}\summ{r=0}{1}{}\summ{s=0}{1}{}K_1\paran{p,q,
r,s}m_L^{p+q} m_{L-1}^{r+s}\\
\end{array}} where

\eq{denk9}{\begin{array}{lcl}
K_1\paran{p,q,r,s}&=&\komb{2}{p}\komb{2}{q}A_{1x}^{2-p}A_{1y}^{2-q}A_{2x}^{1-r}
A_{2y}^{1-s}B_{1x}^{p}B_{1y}^{q}B_{2x}^{r}B_{2y}^{s}f_1\paran{x,y,H_1,H_2}|_{x=0
,y=0}\\
K_2\paran{p,q,r,s,t,v}&=&\komb{2}{r}\komb{2}{s}A_{2x}^{4-(p+r+t)}A_{2y}^{
4-(q+s+v)}B_{2x}^{p+r+t}B_{2y}^{q+s+v}f_2\paran{x,y,H_1,H_2}|_{x=0,y=0}.\\
\end{array}} Here,

\eq{denk10}{\begin{array}{lcl}
A_{km}&=&\cosh{\paran{J_k^z\nabla_m}}\\
B_{km}&=&\sinh{\paran{J_k^z\nabla_m}},\quad k=1,2; m=x,y
\end{array}
} and 
\eq{denk11}{f_n\paran{x,y,H_1,H_2}=
\frac{x+y+H_1+H_2}{X_0^{(n)}}\frac{\sinh{\paran{\beta 
X_0^{(n)}}}}{\cosh{\paran{\beta X_0^{(n)}}}+
\exp{\paran{-2\beta J_1}}\cosh{\paran{\beta Y_0^{(n)}}}}.} 
The effect of the exponential
differential operator to an arbitrary  function $g(x)$ is given by
\eq{denk12}{\exp{\paran{a\nabla_x}}\exp{\paran{b\nabla_x}}g\paran{x,y}=g\paran{
x+a,y+b}} with arbitrary
constants  $a,b$.

These coefficients in nonlinear equation system given in Eq. \re{denk8} and 
defined in Eq. \re{denk9} can be calculated from the definitions given in Eq. 
\re{denk10} and Eq. \re{denk12}. After obtaining these coefficients, one can 
solve the nonlinear equation system given by Eq. \re{denk8} numerically by 
standard methods such as Runge-Kutta methods. The order parameter of the whole 
system is defined by the sum of the magnetizations of each cylinder,
\eq{denk13}{m=\frac{1}{L}\summ{i=1}{L}{m_i}.}

We can construct the hysteresis loops which are nothing but the variation
of the $m$ with the magnetic field by sweeping field $-H\rightarrow H$ and calculation magnetization en each step and then the sweeping field to the reverse direction and again calculating magnetizations. 
The quantity that characterizes the hysteresis loops is hysteresis loop area  (HLA) will be calculated in this work. HLA 
corresponds to the energy loss due to the hysteresis, and it is simply the area of the closed hysteresis loop.

\section{Results and Discussion}\label{results}

In order to obtain dimensionless Hamiltonian parameters let us define
\eq{denk14}{r=\frac{J_1}{J_2}, t=\frac{k_BT}{J_2}, h=\frac{H}{J_2}.
} Also, $\Delta_1=\Delta_2=\Delta$ has been used throughout the calculations.

\subsection{Phase Diagrams}

\begin{figure}[h]\begin{center}
\epsfig{file=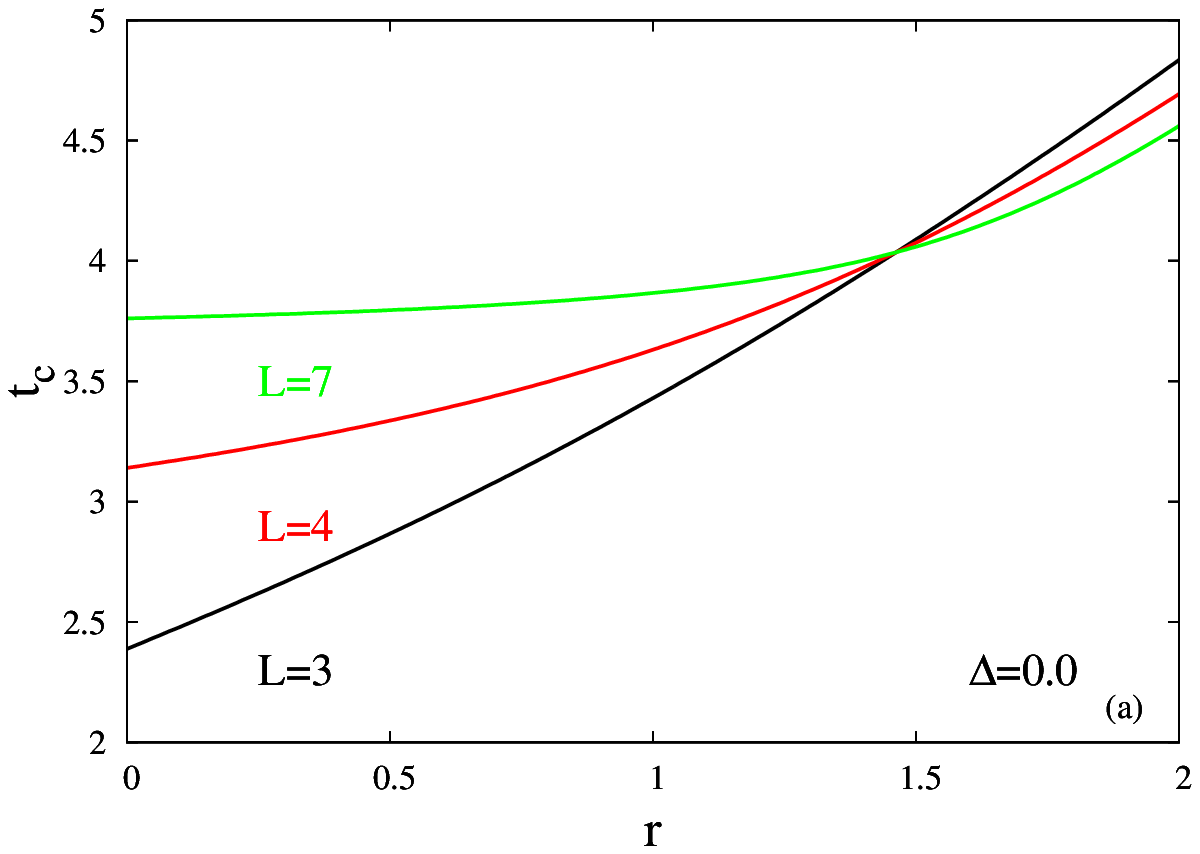, width=7.4cm}
\epsfig{file=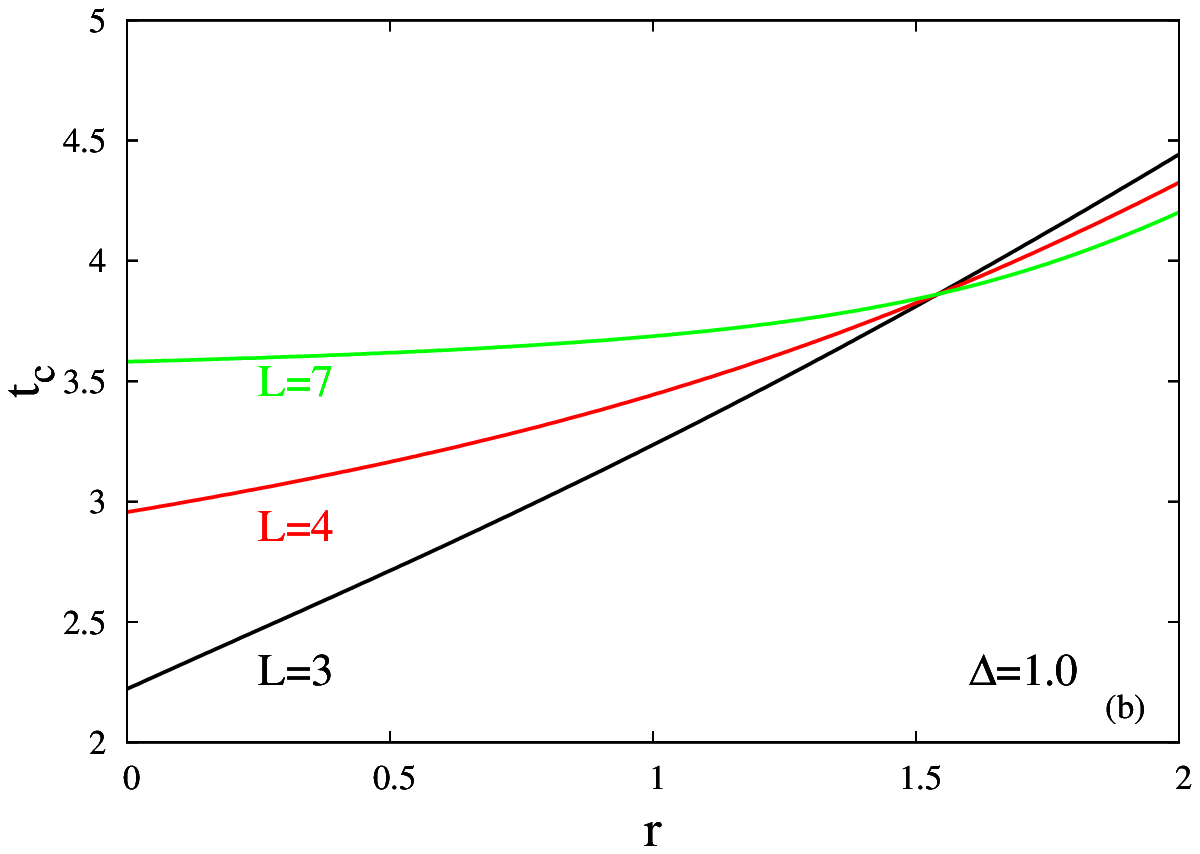, width=7.4cm}
\end{center}
\caption{Phase diagrams of then MWNT that have different thickness in $(t_c,r)$ 
plane for (a) Ising model and (b) isotropic Heisenberg model} 
\label{sek2}\end{figure}

The critical temperature of the system can be obtained from the solutions of the linearized forms of the nonlinear equation system Eq. \re{denk8}. The variation of the critical temperature with $r$ can be seen in Fig. \ref{sek2} for 
different thickness and two limiting cases namely, (a) Ising case and (b) 
isotropic Heisenberg case. The phase diagrams in $(t,r)$ plane intersects at 
one 
point, namely special point. This situation is typical of all layered structures such as thin films \cite{ref20}. The special point makes the critical temperature of the MWNT independent of the thickness, while left of it the critical temperature of the system increases with rising thickness, the reverse is true for the right side of the special point. The temperature coordinate of this special point is indeed the critical temperature of the corresponding bulk system. Note that, the corresponding bulk system is in the hexagonal geometry.

The special point coordinates ($t^*,r^*$) of the Ising model on this MWNT 
geometry found as ($4.039,1.469$) and for the isotropic Heisenberg model it 
is 
($3.864,1.548$). The temperature coordinate of the special point of the Ising 
model ($t^*=4.039$) lies between the critical temperature obtained within the 
same formulation of the square lattice $t_c=3.025$ and simple cubic lattice 
$t_c=5.039$ \cite{ref21}. This is consistent since bulk counterpart of the MWNT 
is hexagonal geometry which has a lower critical temperature than the cubic geometry, due to the fewer number of nearest neighbors of it. 
This value of isotropic Heisenberg case can be compared with the result of the 
simple cubic thin film geometry within the same formulation, which is 
($4.891,1.345$) \cite{ref20}. We can say that the value $t^*=1.548$ is higher 
than the value for cubic isotropic Heisenberg thin film ($t_c=1.345$) with the same 
reasoning as explained above.

We can say that, the special point shifts towards to the right lower regions of 
the  $(t,r)$ plane when the system goes from the Ising model to the isotropic 
Heisenberg model with passing $XXZ$ type anisotropic Heisenberg model.

\subsection{Thermodynamical properties}

\begin{figure}[h]\begin{center}
\epsfig{file=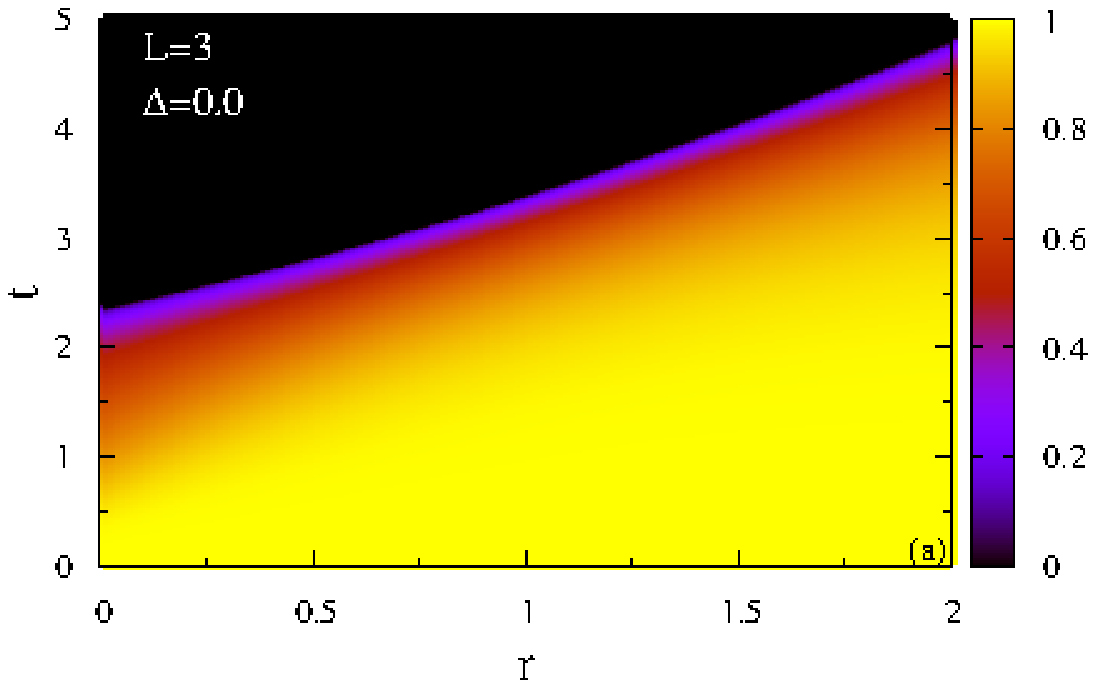, width=7.4cm}
\epsfig{file=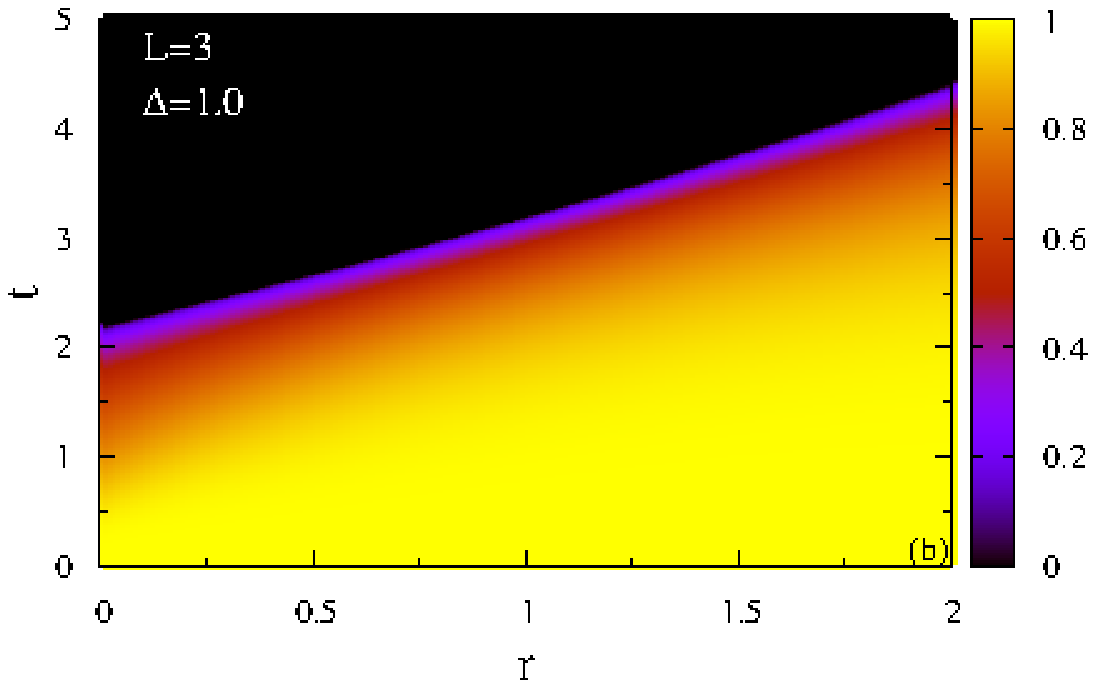, width=7.4cm}

\epsfig{file=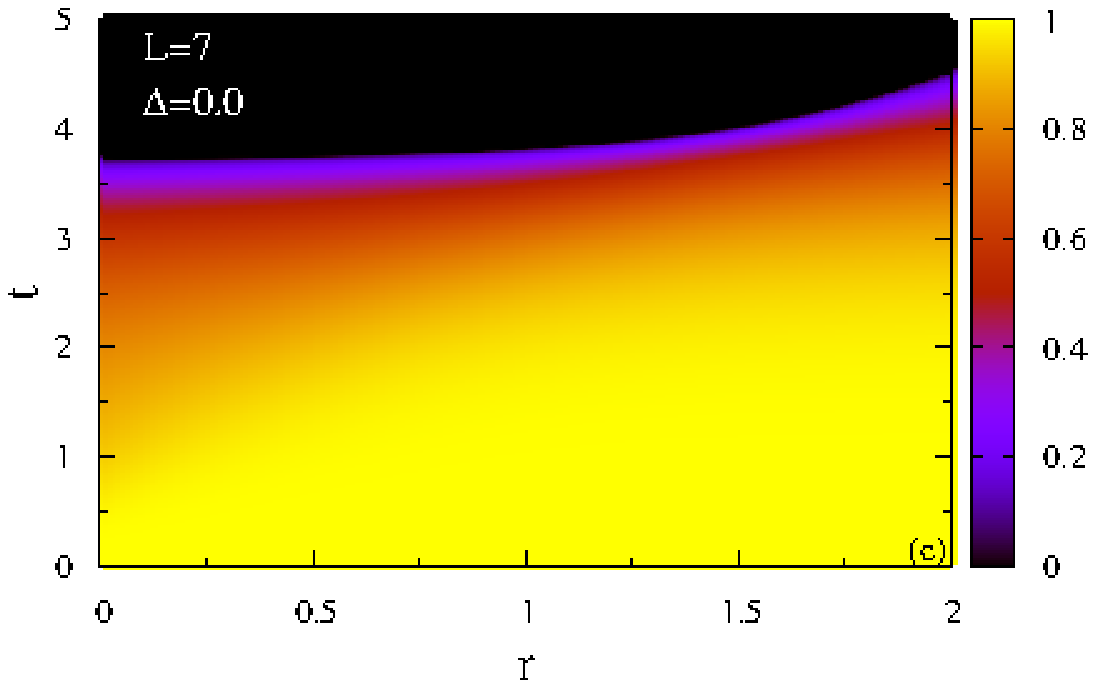, width=7.4cm}
\epsfig{file=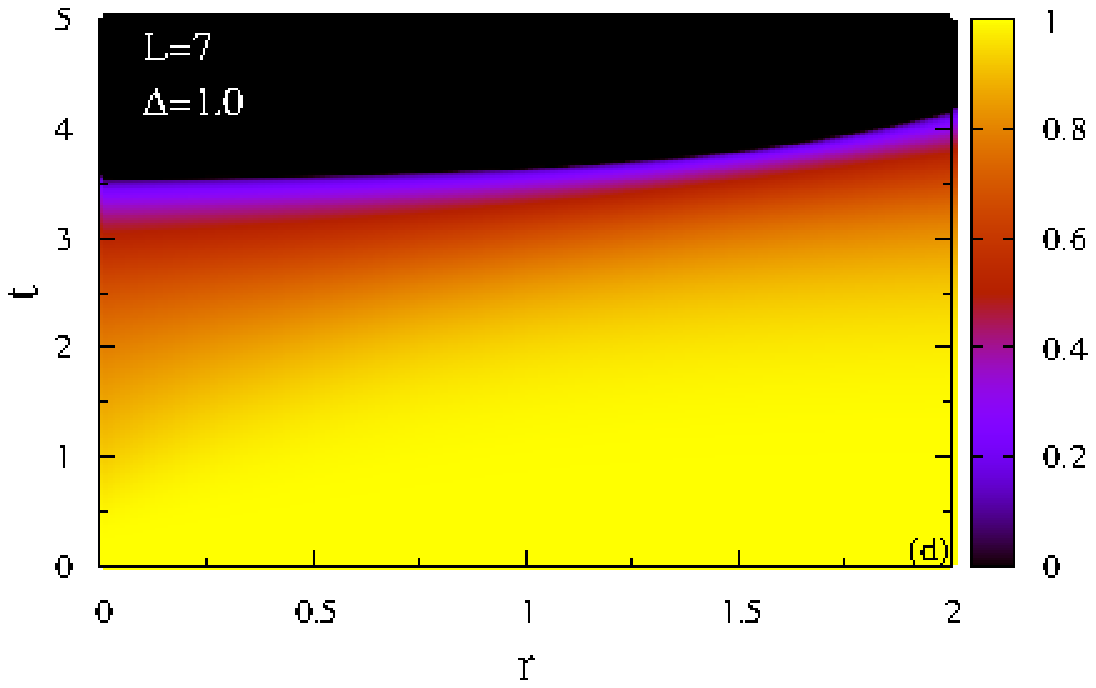, width=7.4cm}
\end{center}
\caption{Variation of magnetization in $(t,r)$ plane of MWNT for selected values 
of 
(a) $L=3, \Delta=0.0$ (a) $L=3, \Delta=1.0$ (a) $L=7, \Delta=0.0$  and (a) $L=7, 
\Delta=1.0$. } 
\label{sek3}\end{figure}

We can see the variation of the magnetization in $(t,r)$ plane as contour 
surface, for selected values of $L$ and 
$\Delta$ in Fig. \re{sek3}. By comparing  Figs. \re{sek3} and \re{sek2} we can 
say that, variation of magnetization is consistent
with the phase diagrams. The magnetization of the system gradually decreases when one move in $(t,r)$ to the phase boundary. After
this boundary, the magnetization of the system is zero. Changing of magnetization by temperature is gradually decreasing behavior as expected. Behavior in $r$ direction depends on the temperature and the other parameters of Hamiltonian. For instance, when 
the temperature is $3.0$, rising $r$ could strengthen the ferromagnetic phase of 
the system (see, for instance Fig. \re{sek3} (c)), or 
induce the phase transition from the paramagnetic phase to the ferromagnetic 
phase (see, for instance Fig. \re{sek3} (a)). In general, we can say that ordered phase covers a smaller area of $(t,r)$ plane for isotropic Heisenberg limit, in comparison with Ising limit
(compare Figs. \re{sek3} (b) with (a) and (d) with (c)). Thus in general, 
decreasing of the anisotropy in the
exchange interaction (means that rising $\Delta$), shrinks the ferromagnetic region in $(t,r)$ plane. This shrinking behavior is more apparent for lower values
of $L$. 

These facts are related to the effect of anisotropy in the exchange interaction. In the 
Ising limit $(\Delta=0)$, model is extremely anisotropic, 
thus more energy needed to destroy the effect of the exchange interaction. This means that a higher critical temperature. In the opposite, isotropic Heisenberg limit $(\Delta=1.0)$ it is more easy to destroy the ferromagnetic order due to the more isotropic
interactions between the spins, in comparing with Ising case.
On the other hand, nanotubes that have more layer have bigger "exchange interaction per spin" due to the number of layers. Then the critical temperature 
is higher than the nanotube that has little 
$L$ value, when the other values of parameters of Hamiltonian are the same.

\begin{figure}[h]\begin{center}
\epsfig{file=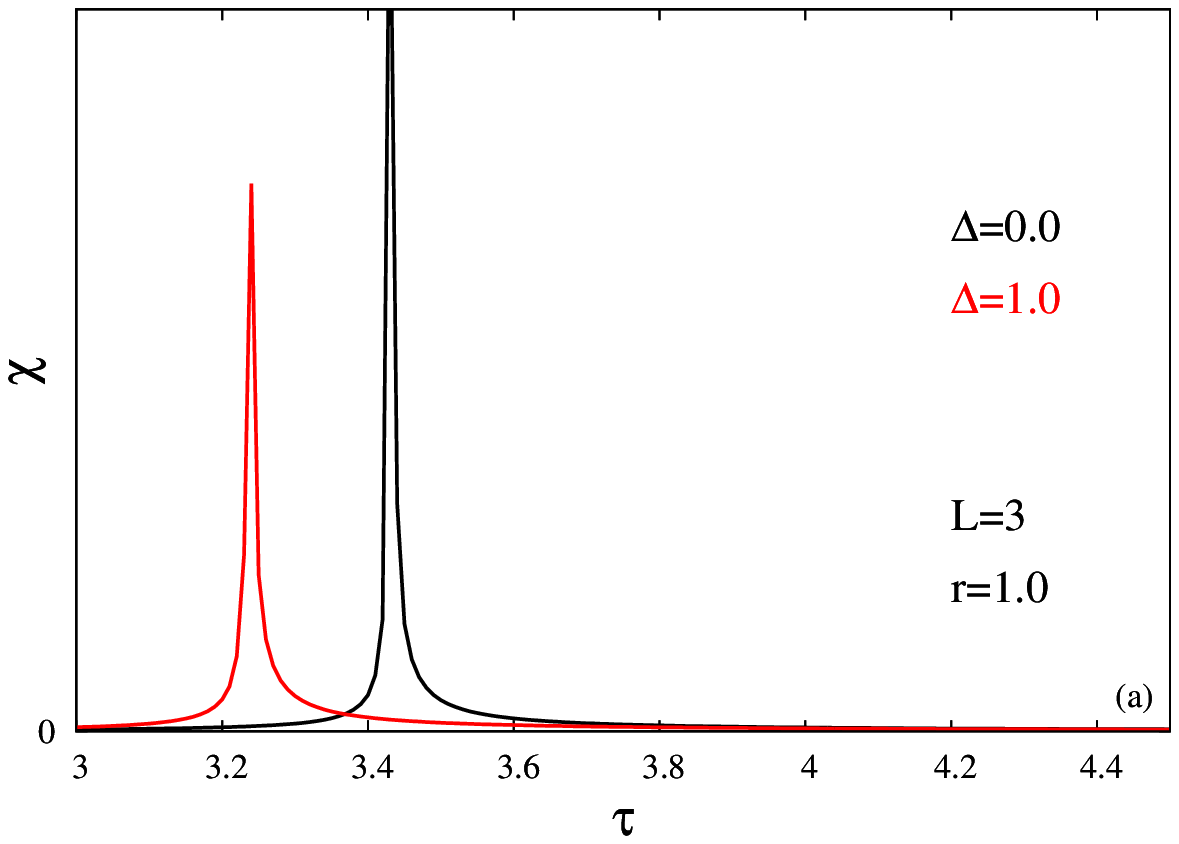, width=7.4cm}
\epsfig{file=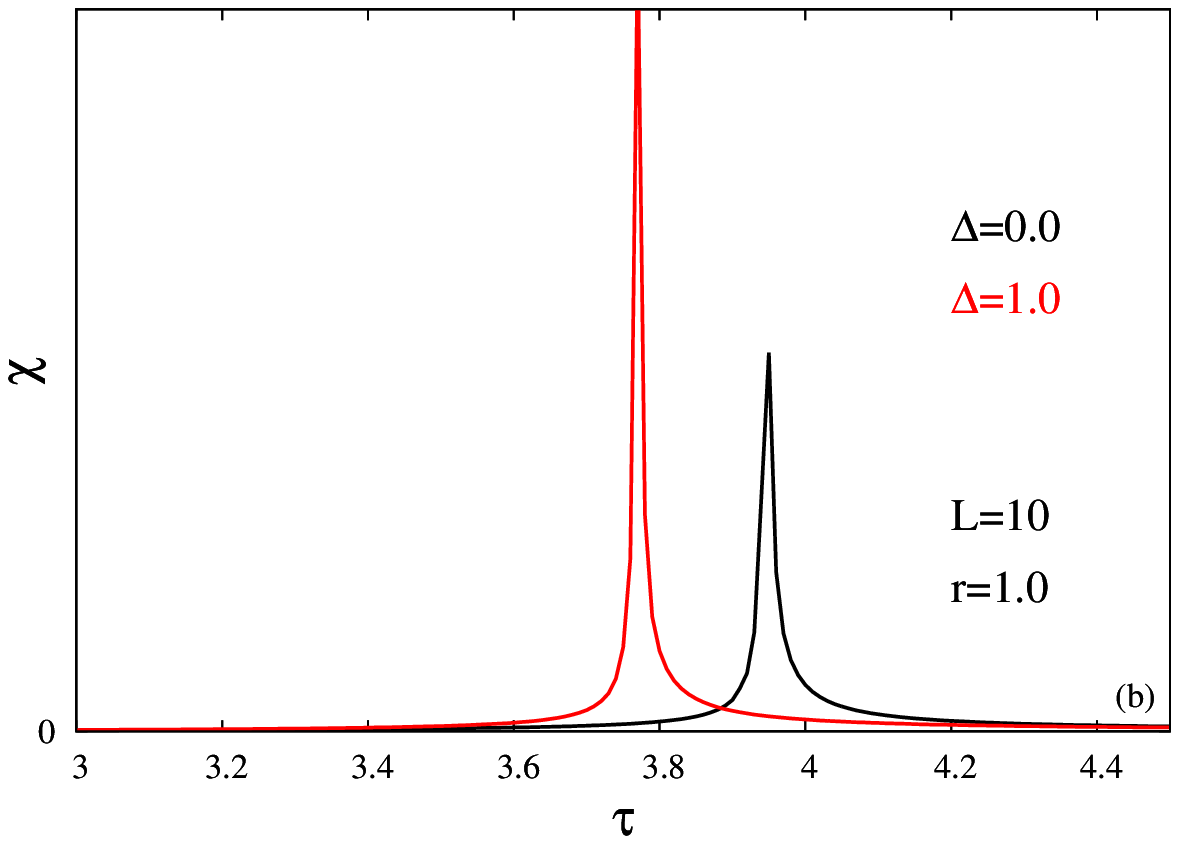, width=7.4cm}
\end{center}\caption{Variation of magnetic susceptibility with different 
thickness of MWNT with temperature
for selected values of  $r=1.0$ (a) $L=3$ and (b) $L=10$. } 
\label{sek4}
\end{figure}

Another important thermodynamical property is magnetic susceptibility which can be measured in experiments. Magnetic susceptibility
is a response of the system by magnetization to the external magnetic field. 
Behaviors of magnetic susceptibilities by changing temperature can be seen
in Fig. \re{sek4}. As seen  Fig. \re{sek4} (a) and (b), rising anisotropy in the exchange interaction causes in the peak of the curve, shifts toward to the right in $(\xi,t)$ plane. This behavior is consistent with the phase diagrams. The peak of the susceptibility
curve occurs at the critical temperature of the system. As seen in Fig. 
\re{sek2}, for the value of $r=1.0$, critical temperature rises
by decreasing $\Delta$, means rising anisotropy in the exchange interaction. 
Besides we can say from Fig. \re{sek4} that, rising
anisotropy in the exchange interaction causes to more intense response to the magnetic field for lower values of the number of layers. This
fact can be seen from the peak values of the susceptibility curves in Fig. 
\re{sek4} (a). When the nanotube has a higher number of layers, this situation getting reverse, i.e., more intense reply comes from the less anisotropic case $(\Delta=1.0)$.

\subsection{Hysteresis properties}

Reply of the system to the magnetic field is indeed history dependent. This causes to hysteresis loops. Typical hysteresis loops of the 
MWNT can be seen in Fig. \re{sek5} for some selected values of parameters of Hamiltonian and temperature.  At first
sight, the effect of the temperature on the hysteresis loops can be seen by comparing Figs. \re{sek5} (a) with (b). Rising temperature
shrinks the loops. This is obvious since rising temperature weakens the order of the system due to the rising thermal fluctuations. On
the other hand, rising $r$ enlarge the loops as seen in Fig. \re{sek5} (compare 
(c), (a) and (d) ). This is due to the strengthened exchange
interaction, which means that strengthened magnetic order of the nanotube. But, 
we cannot say these previous type general conclusions
about the effect of the changing number of layers. For instance, while loop 
related to the $L=3$ lies inside of the 
loop related to the $L=7$ for the values of the parameters of $t=1.9, 
\Delta=1.0, r=1.0$ (see Fig. \re{sek5} (a)), by raising $r$, 
$L=3$ curve lies out of the curve related to the $L=7$ (see Fig. \re{sek5} (d)). 
Moreover, at some values of the parameters, $L=3$ curve
is paramagnetic while $L=7$ is ferro (see Fig. \re{sek5} (b)). 
In conclusion, the variation of the hysteresis loop with the number of the layer depends on the values of the other parameters. 

\begin{figure}[h]\begin{center}
\epsfig{file=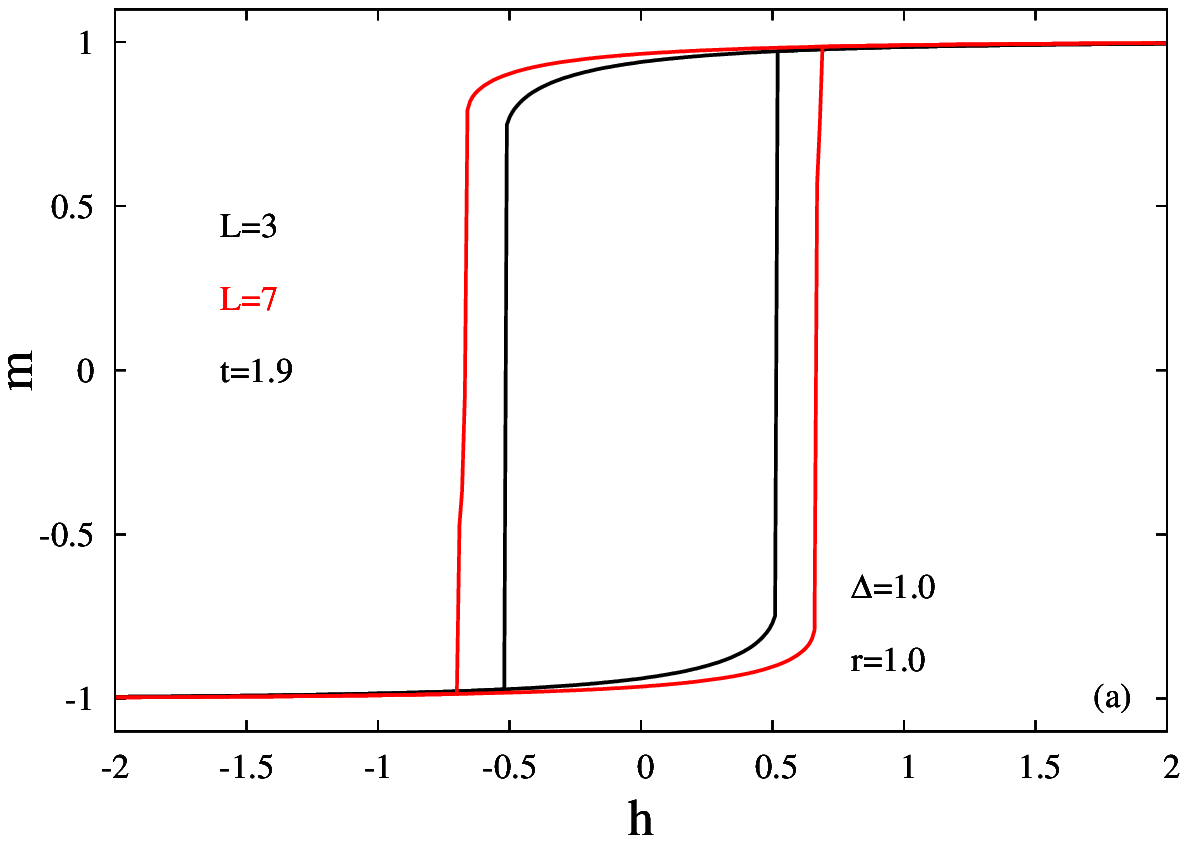, width=7.4cm}
\epsfig{file=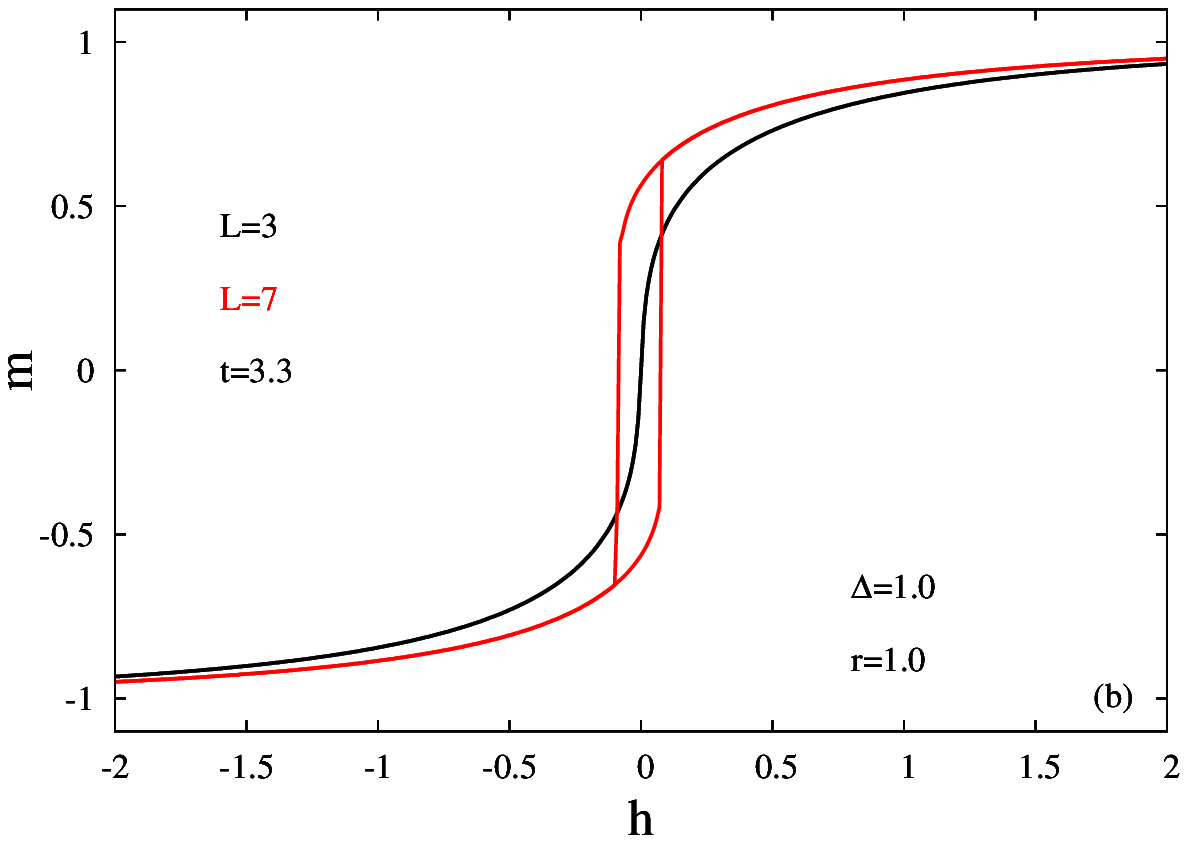, width=7.4cm}

\epsfig{file=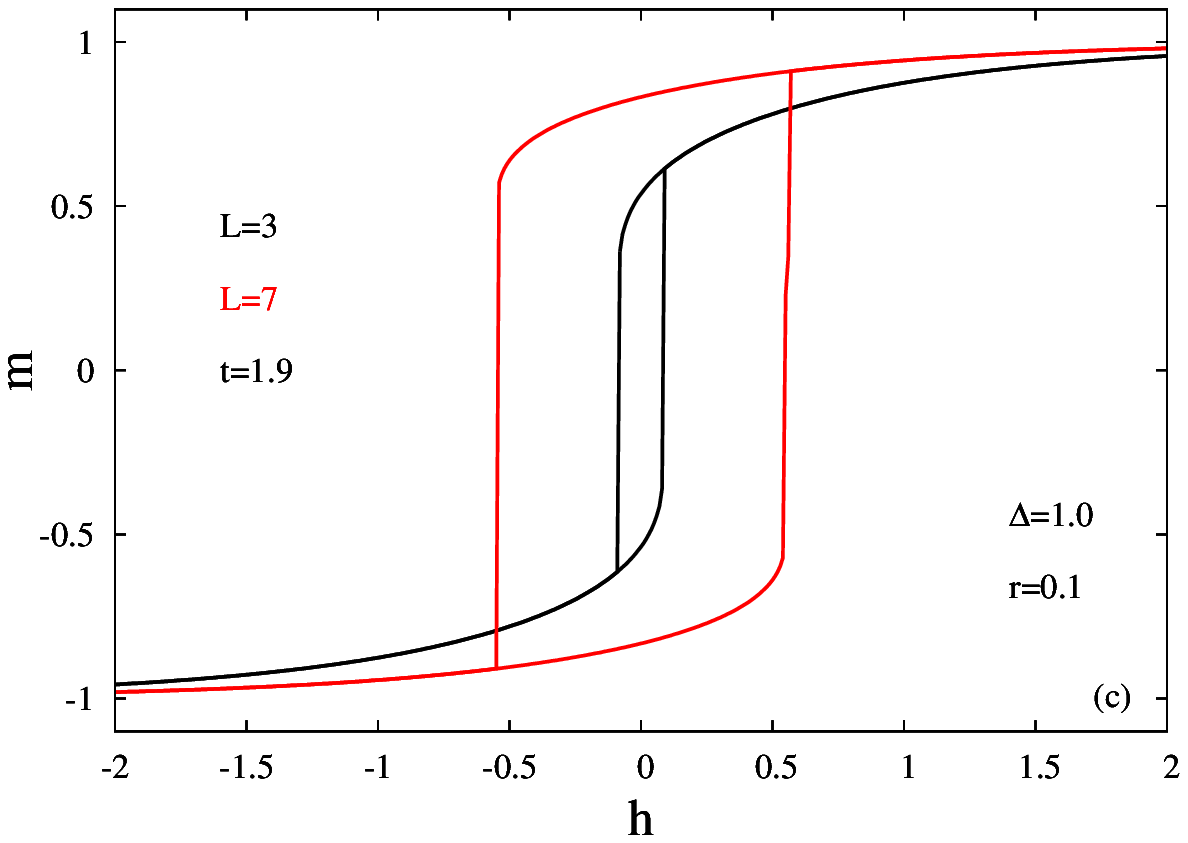, width=7.4cm}
\epsfig{file=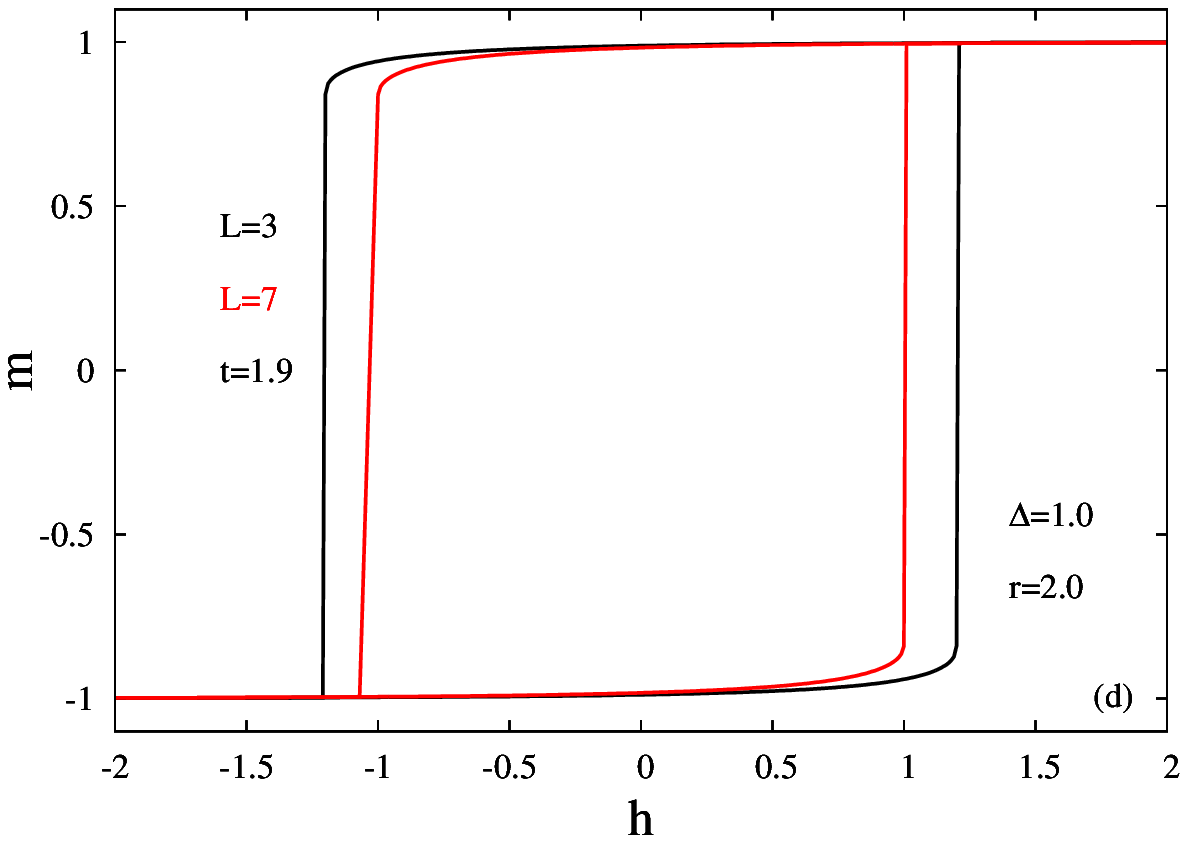, width=7.4cm}
\end{center}
\caption{Selected hysteresis loops of MWNT with $\Delta=1.0$, $L=3$ and $L=7$ 
for the values of 
(a) $t=1.9, r=1.0$ (b) $t=3.3, r=1.0$ (c) $t=1.9, r=0.1$ and (d) 
$t=1.9, r=2.0$. } 
\label{sek5}\end{figure}

One quantity related to the hysteresis loop is hysteresis loop area (HLA) which is nothing but the area covered by the
closed hysteresis loop. HLA is the measure of the energy loss due to the hysteresis. 
The behavior of the HLA in $(t,r)$ plane can be seen in 
Fig. \re{sek6}, for selected values of $L=3,7$ and $\Delta=0.0,1.0$.

\begin{figure}[h]\begin{center}
\epsfig{file=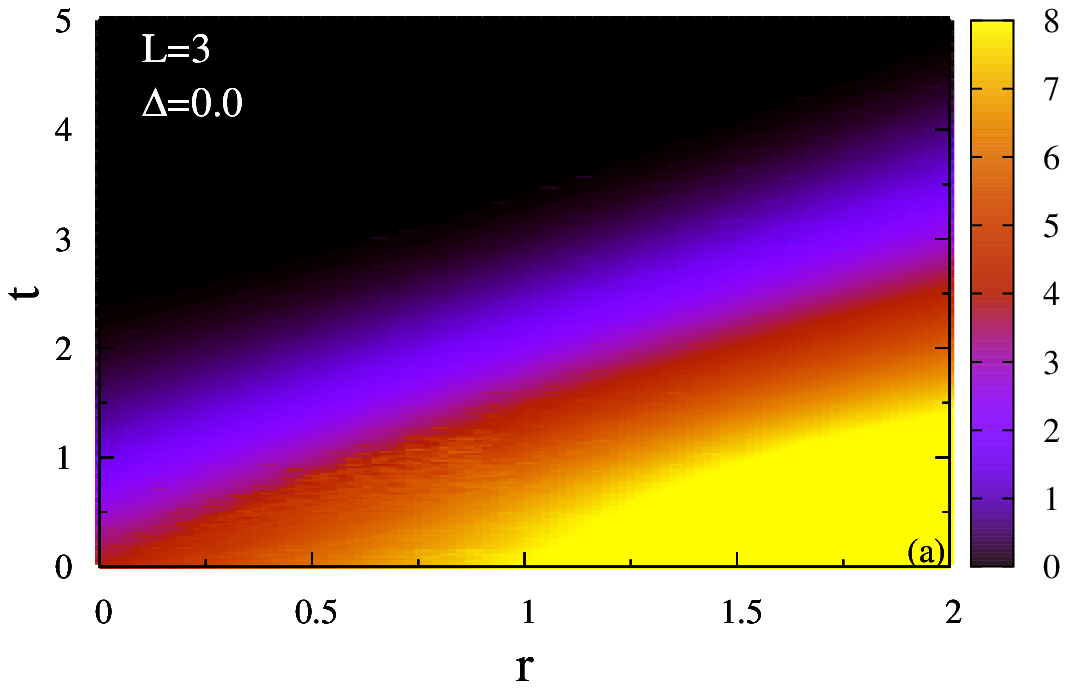, width=7.4cm}
\epsfig{file=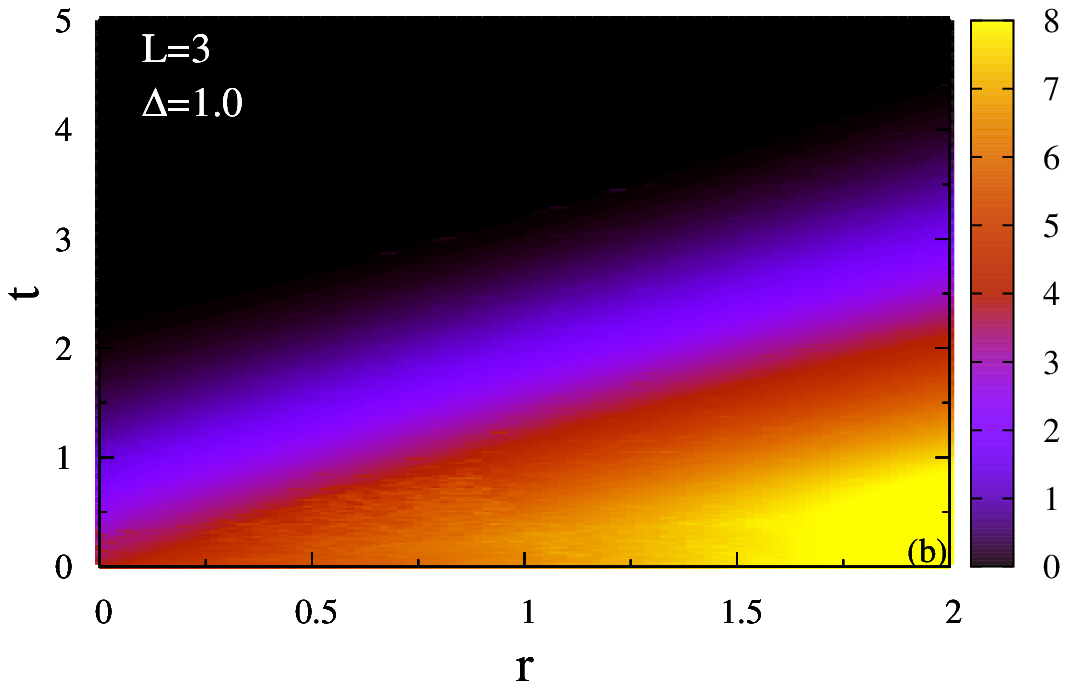, width=7.4cm}

\epsfig{file=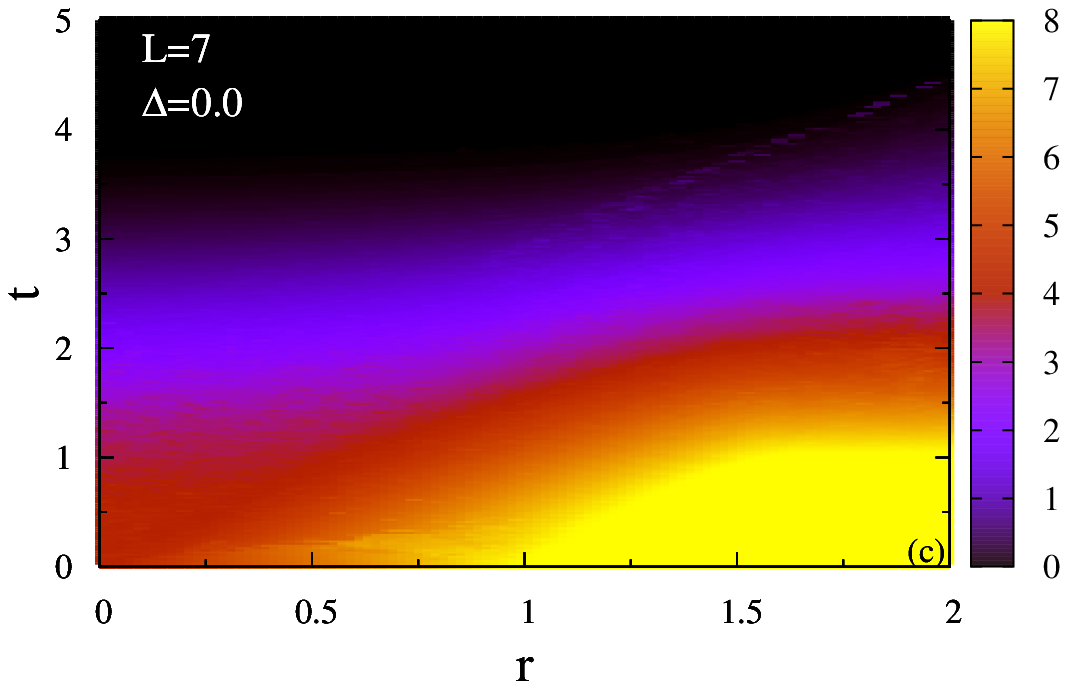, width=7.4cm}
\epsfig{file=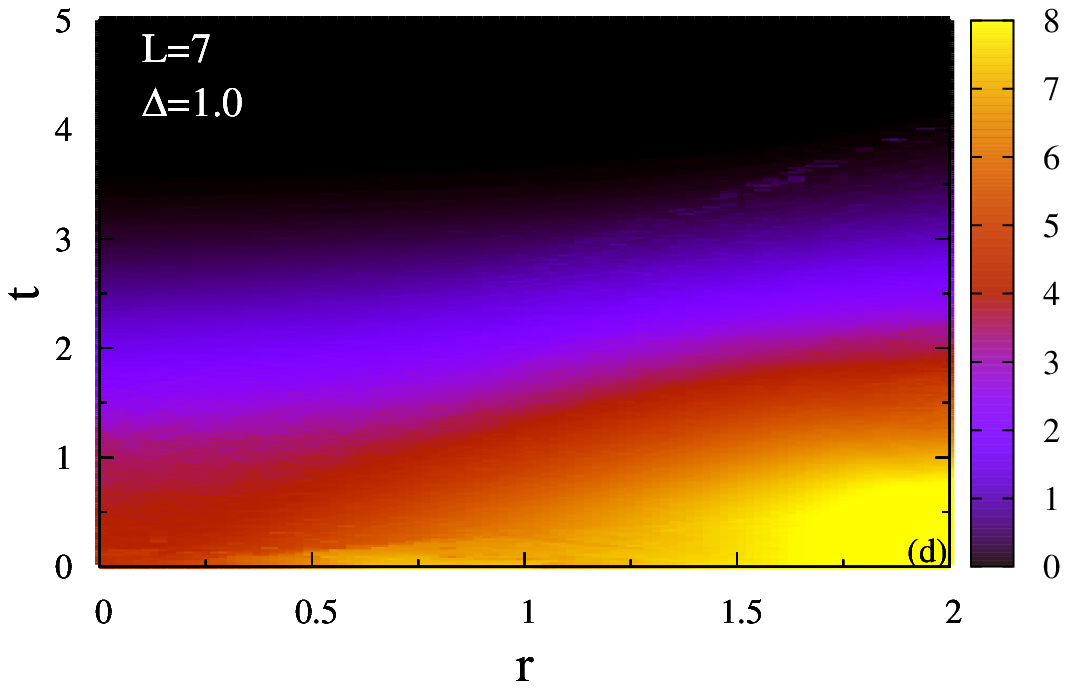, width=7.4cm}
\end{center}
\caption{Variation of HLA in $(t,r)$ plane of MWNT for selected values of 
(a) $L=3, \Delta=0.0$ (b) $L=3, \Delta=1.0$ (c) $L=7, \Delta=0.0$  and (d) $L=7, 
\Delta=1.0$. } 
\label{sek6}\end{figure}

In general, we can say about the relation between the number of layers and HLA, a rising number of layers causes an increment of the HLA.
In other words, thicker nanotubes have higher HLA than the thinner ones. This is valid for any value of anisotropy in the exchange interaction (see Fig. \re{sek5}). One other point is related to the regions that have non zero HLA in $(t,r)$ plane. 
Consistent with Fig. \re{sek2}, thicker nanotubes have a broader region in $(t,r)$ plane which has nonzero HLA.  This result is consistent
by Fig. \re {sek2}, since nonzero HLA corresponds to the ferromagnetic order of the nanotube.

\section{Conclusion}\label{conclusion}

Heisenberg model for the spin-1/2 particles on MWNT geometry has been solved within the formulation EFT-2. Effect of the anisotropy in the exchange interaction and thickness of MWNT on the thermodynamical properties have been obtained. By this way, two limits of the model namely, Ising
model and isotropic Heisenberg model compared on a MWNT geometry. 

Firstly, as in the layered systems such as thin films, phase diagrams of the MWNT displays a special point. At this point (special value of $r^*$), the critical temperature of the system gets independent of the number of layers. For the values that provide 
$r<r^*$, thicker nanotubes have higher critical temperature and vice versa. Remember that, the value of $r$ is the ratio between the exchange interactions of the nanotube. Since exchange interaction is related to the overlapping of the atomic orbitals which constitute the nanotube, we can conclude that effect of the thickness of the MWNT on the critical temperature depends on the constituent atoms of the nanotube. While some MWNTs, thicker nanotubes have a higher critical temperature in comparison by the thinner ones, some other nanotubes which composed of different atoms this relationship may be in
the opposite form. 

Another difference between the MWNT materials is anisotropy in the exchange interaction. 
Rising anisotropy in the exchange interaction means that, changing model from the isotropic Heisenberg model to the Ising model. 
In general critical temperature rises for the same values of the other Hamiltonian parameters. This is also valid for bulk materials.

After analysis of the order parameter, some conclusions about the behavior of the magnetic susceptibility obtained.
We can conclude about the measure of the reply to the magnetic field at the critical temperature that, for lower valued $L$ response to
the magnetic field becomes more intense when the anisotropy in the exchange interaction rises (i.e., $\Delta: 1\rightarrow 0$). 
On the other hand, when the MWNT gets thicker, this relation get reverse, i.e.,   the peak value of the 
MWNT decreases, when the anisotropy in the exchange interaction rises (i.e., $\Delta: 1\rightarrow 0$). 

Last observations are on the hysteresis loops of the MWNT. After detailed investigations on the loops and HLAs, we conclude that relation between the HLA values of the thicker and thinner nanotubes depends on the $r$ value of the system. 

We hope that the results  obtained in this work may be beneficial 
form both theoretical and experimental point of view.

\end{document}